\documentstyle[11pt,preprint,aps]{revtex}
\input epsf
\epsfverbosetrue

\newlength{\oldbaselineskip}

\draft
\preprint{\setlength{\baselineskip}{2.6ex}\hfil 
\vbox{\hbox{hep-ph/9609314\\}\hbox{TRI--PP--96--49}
\hbox{September 1996}}}

\title{Supersymmetric Time Reversal Violation in Semileptonic Decays 
of Charged Mesons}
 
\author{Guo-Hong Wu \footnote{\footnotesize gwu@alph02.triumf.ca
\vspace{-.15in}} 
 and  John N. Ng \footnote{\footnotesize misery@triumf.ca}}

\address{TRIUMF Theory Group\\4004 Wesbrook Mall, Vancouver, B.C., 
Canada V6T 2A3}

\begin{document}
\setlength{\baselineskip}{24pt}
\setlength{\oldbaselineskip}{\baselineskip}
\maketitle

\vskip -.3in
\begin{abstract}

  We provide a general analysis of time reversal violation arising
from misalignment between quark and squark mass
eigenstates. In particular, we focus on the possibility of large
enhancement effects due to the top quark mass.
For semileptonic decays of the charged mesons, 
$K^+ \rightarrow \pi^0 \mu^+ \nu_{\mu}$,
$D^+  \rightarrow \overline{K}^0 \mu^+ \nu_{\mu}$,
and $B^+ \rightarrow \overline{D}^0 \tau^+ \nu_{\tau}$, the transverse
polarization of the lepton $P^{\bot}_l$ is a $T$-odd observable that is of 
great experimental interest. 
It is noted that under favorable choice of parameters, 
$P^{\bot}_{\mu}$ in $K^+_{\mu3}$ decay can be detectable
at the ongoing KEK experiment and it holds a promising prospect for 
discovery at the proposed BNL experiment. Furthermore,
$P^{\bot}_{\tau}$ in $B^{\pm}$ decay could well be within the reach of 
$B$ factories, 
but  $P^{\bot}_{\mu}$ in $D^{\pm}$ decay is not large enough 
for detection at the  proposed $\tau$-charm factory. 
\end{abstract} 

\setlength{\baselineskip}{\oldbaselineskip}
\newpage
 
 In supersymmetric (SUSY) theories,
different unitary transformations are generally required for the
quark and squark gauge eigenstates to reach their respective 
mass eigenstates \cite{fcnc}. The difference between the quark and squark 
transformations is referred to as quark-squark misalignment (QSM), 
a consequence of which is the appearance of nontrivial family 
mixing matrices in the gluino--quark--squark couplings and thus 
new contributions to flavor changing neutral current (FCNC) processes
arise.
The physical phases in these mixing matrices could  serve as 
new sources of hadronic $CP$ and $T$ violation 
(assuming $CPT$ invariance),
and a similar mechanism can operate in the lepton sector. 
 In this context, the electric dipole moments of the neutron and the 
electron have recently been studied in the 
minimal supersymmetric standard model (MSSM) \cite{BV},
the minimal SUSY $SO(10)$ \cite{hall}, 
and SUSY grand unified models with an intermediate scale \cite{desh}.  

   In this letter, we study a different manifestation of 
QSM, i.e. its effects on the $T$-odd transverse lepton polarization 
 in semileptonic decays of charged mesons,
 $K^+ \rightarrow \pi^0 \mu^+ \nu_{\mu}$,
$D^+ \rightarrow \overline{K}^0 \mu^+ \nu_{\mu}$,
and $B^+ \rightarrow \overline{D}^0 \tau^+ \nu_{\tau}$.
This  observable measures the $T$-odd triple correlation
among the spin of the charged lepton and the momenta of two final state 
particles. In the rest frame of the decay meson, it is defined as 
\cite{Cabibo} 
\begin{eqnarray}
P^{\bot}_l & \equiv & \frac{ \bf{\hat{s}}_l 
\cdot (\bf{n}_M \times \bf{n}_l )}{
|\bf{n}_M \times \bf{n}_l|},
\end{eqnarray}
where $\bf{\hat{s}}_l$ is the
unit vector along the the charged lepton  spin direction,  
and $\bf{n}_M$ and 
$\bf{n}_l$ are unit vectors along the three momenta
of the final state meson  and the charged lepton respectively.
For the charge conjugate processes of the above decays, $CP$-violating
interactions contribute to $P^{\bot}_l$ with an opposite sign \cite{OK}.

A nonzero $P^{\bot}_l$ arises from the interference between two 
amplitudes with different phases. In this case it involves 
the standard model weak decay amplitude 
 and an effective scalar amplitude \cite{mich}.
As there is only one charged particle in the final state,
the electromagnetic final state interaction contribution 
to $P^{\bot}_l$ is expected to be small,
of order $10^{-6}$ for $K^+_{\mu3}$ decay \cite{zhit}.
The standard model $T$-violation effect is vanishingly small \cite{GV},
 leaving enough room for new physics to come in. 
For example, in the Weinberg model of $CP$ violation \cite{wei},
where $P^{\bot}_l$ arises from the interference
between two tree level amplitudes, 
 an effect as large as $10^{-3}$ can be obtained
for the transverse muon polarization in $K^+_{\mu3}$ decay \cite{pretv},
and much larger contributions are possible for the transverse $\tau$
polarization in $B^+ \rightarrow \overline{D}^0 \tau^+ \nu_{\tau}$
\cite{AES,garisto,Grossman}. 
In the MSSM, if one assumes that $T$ violation comes from the complex 
soft SUSY breaking operators and neglects squark family mixing,
 $P^{\bot}_l$ will be suppressed by the strange quark mass
 $m_s$ for $K^{+}_{\mu3}$, the charm quark mass $m_c$ for $D^+_{\mu3}$, 
and the $b$ quark mass $m_b$ for $B^+_{\tau3}$,  
 and would be too small to to be seen~\cite{cf}.

  With the squark family mixing taken into account,  $P^{\bot}_l$
in the semileptonic decays of charged  kaon, D and B mesons 
could become sensitive to quarks of the third 
family, in particular to the top quark. This comes about in two ways.
Firstly, both the $\tilde{t}_L$--$\tilde{t}_R$ stop mixing 
and the charged Higgs coupling to $\tilde{t}_R$ have strengths
proportional to the top mass $m_t$.
Secondly, due to the renormalization group equation (RGE) running
from a higher scale, 
the third generation squarks tend to be lighter at the weak scale
than squarks of the first two generations because of the large top Yukawa 
coupling. 
These lead to the  possibility that the $\tilde{t}$ and $\tilde{b}$
squarks could give the largest contribution to  $P^{\bot}_l$.
This possibility will be the focus of this letter.

 The various squark family mixing matrices can be determined 
in specific models,
however for the sake of generality we will not be restricted
 to a particular one but rather give a general description.
We refer interested readers to the literature \cite{umix,uhall} for
a discussion on models.
 The mixing matrices characterizing the relative rotations in flavor 
space between the four types of squarks ($\tilde{u}_L$, $\tilde{u}_R$,  
$\tilde{d}_L$ and $\tilde{d}_R$) and their corresponding quarks 
are denoted by $V^{U_L}$, $V^{U_R}$, $V^{D_L}$, and $V^{D_R}$ respectively. 
These matrices appear in the quark-squark-gluino couplings, and are
constrained by FCNC processes \cite{fcnc}. 
These constraints can be divided into two classes.
Firstly, products of $V^D$'s are bounded by FCNC in the down quark sector,
using $K\overline{K}$ and $B\overline{B}$ mixing, 
$b\rightarrow s \gamma$, $\epsilon_K$, and $\epsilon^{\prime}_K$.
Secondly, products of $V^U$'s are bounded by FCNC  in the up quark sector,
and $D\overline{D}$ mixing provides the only available constraint.
Note that $D\overline{D}$ mixing can put nontrivial bounds on
products of $V^{U_{L,R}}_{32}$ and $V^{U_{L,R}}_{31}$, but not on the 
individual matrix element. Therefore large mixing in 32 or 31 up type
 squark families is not ruled out \cite{worah}.
We also note that the neutron electric dipole moment can constrain 
 a particular combination of phases and product of two different squark 
generational mixings, but not the individual mixing matrix element.
In contrast to FCNC processes,  SUSY contributions to
charged current processes, including the semileptonic decays of
 charged mesons, can involve products of $V^U$ and $V^D$ that are not
 subject to direct constraints from FCNC.
By invoking the mixing matrix $V^{U_R}$ in the loop, 
 these charged meson decays can be  enhanced by the top quark mass.
 The usual KM matrix is denoted as $V^{KM}$.

The physical phases that are responsible for $T$-violation
can arise from the squark family mixing matrices and 
other soft SUSY breaking operators including the gaugino masses and the
$A$ terms.
Typically, $T$-odd effects involve the difference of two or more 
physical phases. 
For simplicity of illustration of  the physics and for
  an estimate of the $T$-odd effects, only the phases from the
squark mixing matrices will be kept.
Throughout this letter, we will neglect the mixing between the left and
right squarks except for the top squarks,
as we expect them to play a more important role.
Also we will employ an insertion approximation for the 
$\tilde{t}_L$--$\tilde{t}_R$ mixing where appropriate.

   The dominant contribution to  $P^{\bot}_l$ is thus expected to come
from the gluino-$\tilde{t}$-$\tilde{b}$  loop diagrams with $W$ 
and charged Higgs boson exchange (figures 1 and 2).
To linear order in the external momenta and to first order in the
$\tilde{t}_L$--$\tilde{t}_R$ mixing,
the $W$ exchange diagram (fig.~1) for kaon decay can be evaluated to give 
\begin{eqnarray}
{\cal L}_{eff,W}  &= & [C_W^{+} ( p+p^{\prime})^{\alpha}  
                  +  C_W^{-} ( p-p^{\prime})^{\alpha}]
      (\overline{s}_L u_R)(\overline{\nu}_L \gamma_{\alpha}\mu_L) 
         + h.c., \label{eq:W}
\end{eqnarray}
where $p$ and $p^{\prime}$ are the momenta of the $s$  and $u$ quarks
respectively, and $C_W^{\pm}$ are given by
\begin{eqnarray}
C_W^{\pm} & = &  \frac{1}{18} \frac{\alpha_s}{\pi} \sqrt{2}G_F 
\frac{m_t(A_t-\mu\cot\beta)}{m_{\tilde{g}}^3} {V^{SKM}_{33}}^* 
{V^{D_L}_{32}}^* V^{U_R}_{31} I_W^{\pm},
\end{eqnarray}
where $G_F$ is the Fermi constant, $\alpha_s$ is the QCD coupling 
evaluated at the mass scale of the sparticles in the loop,
$A_t$ is the soft SUSY breaking $A$ term for the top squarks,
$\mu$ denotes the two Higgs superfields mixing parameter, $\tan \beta$ is
the ratio of the two Higgs VEVs, $m_{\tilde{g}}$ is the mass of the  gluino, 
and $V^{SKM}_{ij}$ is the super KM matrix associated with the 
$W$-squark coupling $W^+ {\tilde{u_i}_L}^* \tilde{d_j}_L$. 
Assuming $m_{\tilde{t}_L}=m_{\tilde{t}_R}=m_{\tilde{t}}$ 
for the mass parameters of the left and right top squarks 
and denoting the bottom squark mass by $m_{\tilde{b}}$,
the integrals $I_W^+$ and $I_W^-$ can be written as
\begin{eqnarray}
 I_W^+& = & \int_0^1 dx \int_0^{1-x} dy \frac{24x(1-x-y)}{[
\frac{m_{\tilde{t}}^2}{m_{\tilde{g}}^2}x+
\frac{m_{\tilde{b}}^2}{m_{\tilde{g}}^2}y+
(1-x-y)]^2}  \\
 I_W^-& = & \int_0^1 dx \int_0^{1-x} dy \frac{24x(x-y)}{[
\frac{m_{\tilde{t}}^2}{m_{\tilde{g}}^2}x+
\frac{m_{\tilde{b}}^2}{m_{\tilde{g}}^2}y+
(1-x-y)]^2} \label{eq:IW-}.
\end{eqnarray}
Both integrals $I^{\pm}_W$ are equal to one at 
$\frac{m_{\tilde{t}}}{m_{\tilde{g}}}=\frac{m_{\tilde{b}}}{m_{\tilde{g}}}=1$,
and both increase as $\frac{m_{\tilde{t}}}{m_{\tilde{g}}}$
and/or $\frac{m_{\tilde{b}}}{m_{\tilde{g}}}$ decreases from one.
For example, $I^-_W \simeq 7.9$ when $\frac{m_{\tilde{t}}}{m_{\tilde{g}}}
=\frac{m_{\tilde{b}}}{m_{\tilde{g}}}=\frac{1}{2}$.

 Notice that the $C_W^{-}$ term in eq.~(\ref{eq:W}) 
can be rewritten as an effective scalar
interaction by use of the Dirac equation for the external leptons. 
The term $(p+p^{\prime})^{\alpha} (\overline{s}_Lu_R)$ in eq.~(\ref{eq:W})
 can be Gordon decomposed into a tensor piece and a vector piece.  
The tensor piece can be neglected as the tensor form factor is small.
The vector piece, having the same structure as the standard model
interaction, is readily discarded as its contribution to
$P^{\bot}_l$ vanishes \cite{mich}.
Therefore, $T$-violating effects arise dominantly 
from the interference between the effective scalar  amplitude of the
$C_W^{-}$ term in eq.~(\ref{eq:W}) and the effective
current-current amplitude of the standard model. 

To proceed further, we parameterize the hadronic matrix elements of 
the  quark vector current and quark scalar density  between a kaon and 
a pion state by two form factors
\begin{eqnarray}
\langle \pi^0|\overline{s} \gamma_{\mu} u |K^+ \rangle  & = & 
f_+^{K} (p_K + p_{\pi})_{\mu} + f_-^K (p_K - p_{\pi})_{\mu} \label{eq:Kff} \\
 \langle \pi^0|\overline{s}  u |K^+ \rangle  &\simeq & - 
\frac{f_+^K m_K^2}{m_s} \label{eq:matks} ,
 \end{eqnarray}
where $p_K$ and $p_{\pi}$ denote the kaon and pion momenta, and where
the two form factors $f_+^K$ and $f_-^K$ are functions of 
$(p_K-p_{\pi})^2$. 
Experimentally, $f_-^K$ is small compared to $f_+^K$, 
and its contribution to the scalar matrix element in eq.~(\ref{eq:matks}) 
has been neglected to a first approximation.

The lepton transverse polarization can now be
estimated  following the standard procedure \cite{pretv}.
In the kaon rest frame, for the outgoing muon and neutrino coming at 
right angle for which $P^{\bot}_{\mu}$ is large,
the $W$-exchange contribution  gives
\begin{eqnarray}
P^{\bot}_{\mu}|_{\bf{n}_{\mu}\cdot \bf{n}_{\nu}=0}^W
 & \simeq &  (2\sqrt{2}G_F\sin\theta_c)^{-1} \frac{m_{\mu}m_K}{m_s}
\frac{|\bf{p}_{\mu}|}{E_{\mu}} Im C_W^-  \label{eq:KW}  \\ \nonumber
 & \simeq  &  \frac{\alpha_s}{36 \pi} I_W^{-} 
\frac{|\bf{p}_{\mu}|}{E_{\mu}} \frac{m_{\mu}m_K}{m_s} 
\frac{m_t(A_t-\mu\cot\beta)}{m_{\tilde{g}}^3}
 \frac{Im [V^{SKM*}_{33} V^{D_L*}_{32} V^{U_R}_{31}]}{\sin\theta_c}
\\ \nonumber
 & \simeq &  3 \times 10^{-6} I^-_W 
\frac{|\bf{p}_{\mu}|}{E_{\mu}}
 \frac{m_t(A_t-\mu\cot\beta)}
{m_{\tilde{g}}^2} \frac{100 \; \mbox{GeV}}{m_{\tilde{g}}}
 \frac{Im [V^{SKM*}_{33} V^{D_L*}_{32} V^{U_R}_{31}]}{\sin\theta_c} 
\end{eqnarray}
where we take $\alpha_s \simeq 0.1$, 
$\theta_c$ is the Cabibo angle, $m_{\mu}$, $m_K$ and 
$m_s \simeq 150 \; \mbox{MeV}$ denote the
masses of the muon, the charged kaon and the strange quark, 
and where $\bf{p}_{\mu}$ and $E_{\mu}$ are the outgoing muon 
momentum and energy. 

It is readily seen from eq.~(\ref{eq:KW}) that the actual size of 
$P^{\bot}_{\mu}$ from $W$ exchange is model dependent in two aspects.
Firstly, it depends on the mass spectrum of the SUSY particles,
in particular on $m_{\tilde{g}}$ and on $I^-_W$ through 
$m_{\tilde{t}}/m_{\tilde{g}}$ and $m_{\tilde{b}}/m_{\tilde{g}}$. 
As noted before from eq.~(\ref{eq:IW-}), 
 $P^{\bot}_{\mu}$ can be enhanced by one
order of magnitude if $m_{\tilde{t}}$ and $m_{\tilde{b}}$ decrease from
the mass of the gluino to half of its mass.
Secondly, it is dependent on the mixing matrices $V^{SKM}_{33}$,
$V^{D_L}_{32}$ and  $V^{U_R}_{31}$.
As these parameters are strictly unknown and the bounds on them are model
dependent, the number in eq.~(\ref{eq:KW}) should be taken as qualitative.
For an estimate of the effect, the dimensionful SUSY parameters appearing
in eq.~(\ref{eq:KW}) can be set to be 100 GeV.
With these caveats, we conclude that the magnitude of $P^{\bot}_{\mu}$ 
from $W$ exchange is no larger than a few $\times 10^{-5}$.

 We now consider the contribution from charged Higgs exchange which
  can involve several diagrams.
As we are interested in the largest possible effect
from the top quark, it can be seen that the dominant contribution
involves the  $H^- \tilde{t}_R {\tilde{b}_L}^*$ coupling and a gluino
in the loop (fig.~2).
The effective interaction obtained by integrating out the $\tilde{t}_R$,
$\tilde{b}_L$, $\tilde{g}$ and $H^+$ is given by 
\begin{eqnarray}
{\cal L}_{eff,H^+} & = & 
\frac{C_{H^+}}{m_{H^+}^2} (\overline{s}_L u_R)(\overline{\nu}_L \mu_R)
 + h.c. ,
\end{eqnarray}
where $m_{H^+}$ is the mass of the charged Higgs boson, and $C_{H^+}$ 
is defined as   
\begin{eqnarray}
C_{H^+}& = & -  \frac{2}{3} \frac{\alpha_s}{\pi}
\sqrt{2} G_F m_t m_{\mu} \tan \beta 
\frac{\mu + A_t\cot\beta}{m_{\tilde{g}}} 
{V^{H^+}_{33}}^* {V^{D_L}_{32}}^* V^{U_R}_{31} I_{H^+}, 
\end{eqnarray}
where $V^{H^+}_{ij}$ is the mixing matrix in the charged-Higgs-squark 
coupling $H^+ {\tilde{u_i}_R}^* \tilde{d_j}_L$, 
and the integral function $I_{H^+}$ is given by
\begin{eqnarray}
 I_{H^+} & = & \int_0^1 dx \int_0^{1-x} dy \frac{2}{
\frac{m_{\tilde{t}}^2}{m_{\tilde{g}}^2}x+
\frac{m_{\tilde{b}}^2}{m_{\tilde{g}}^2}y+
(1-x-y)}. 
\end{eqnarray}
which is equal to one at $m_{\tilde{t}}=m_{\tilde{b}}=m_{\tilde{g}}$.
In the above integral $m_{\tilde{t}}$ and $m_{\tilde{b}}$ 
denote the mass parameters of $\tilde{t}_R$ and $\tilde{b}_L$.
The size of $I_{H^+}$ increases as $m_{\tilde{t}}/m_{\tilde{g}}$ and/or
$m_{\tilde{b}}/m_{\tilde{g}}$ decreases, but not as rapidly as
the integral function $I^-_W$. For example, 
$I_{H^+}$ increases to 2.3 as $m_{\tilde{t}}$ and $m_{\tilde{b}}$ decrease
to half the gluino mass.

  In the kaon rest frame and for $\bf{n}_{\mu}\cdot \bf{n}_{\nu}=0$,  
the muon transverse polarization from 
charged Higgs exchange can be estimated as 
\begin{eqnarray}
P^{\bot}_{\mu}|_{\bf{n}_{\mu}\cdot \bf{n}_{\nu}=0}^{H^+} & \simeq &
 - (2\sqrt{2}G_F m_{H^+}^2 \sin\theta_c)^{-1} \frac{m_K}{m_s}
\frac{|\bf{p}_{\mu}|}{E_{\mu}} Im C_{H^+} \label{eq:KH} \\ 
 & \simeq &  \frac{\alpha_s}{3 \pi} I_{H^+} \frac{|\bf{p}_{\mu}|}{E_{\mu}} 
\frac{m_K}{m_s} \frac{m_t m_{\mu} }{m_{H^+}^2} \tan \beta
\frac{(\mu + A_t\cot\beta)}{m_{\tilde{g}}}
\frac{Im [{V^{H^+}_{33}}^* {V^{D_L}_{32}}^* V^{U_R}_{31}]}{\sin \theta_c}
 \nonumber \\
& \simeq &  7\times 10^{-5} I_{H^+} 
\frac{|\bf{p}_{\mu}|}{E_{\mu}} \tan \beta 
\frac{\mu + A_t \cot \beta}{m_{\tilde{g}}}
\frac{(100 \;\mbox{GeV})^2}{m_{H^+}^2} 
\frac{Im [{V^{H^+}_{33}}^* {V^{D_L}_{32}}^* V^{U_R}_{31}]}{\sin \theta_c},
\nonumber
\end{eqnarray}
where we use $\alpha_s\simeq 0.1$ and $m_t=180 \; \mbox{GeV}$.
Note that there are two crucial differences between the Higgs contribution
 of eq.~(\ref{eq:KH}) and the $W$ contribution of eq.~(\ref{eq:KW}). 
Firstly, due to the non-derivative coupling of the 
charged Higgs to the squarks (in contrast to the derivative coupling
for the $W$-squark-squark vertex), the numerical coefficient in
eq.~(\ref{eq:KH}) is one order of magnitude bigger than that of 
eq.~(\ref{eq:KW}).
Secondly, unlike the $W$ exchange, charged Higgs coupling to 
$\overline{\nu}\mu$ is proportional to $\tan  \beta$, and its contribution to
$P^{\bot}_{\mu}$ can be greatly enhanced when $\tan \beta$ is large.
To date the best limit on $\tan \beta$ and $m_{H^+}$ comes from
$b \rightarrow c \tau \nu$ and is given by 
$\frac{\tan \beta}{m_{H^+}} < 0.52 \; \mbox{GeV}^{-1}$ \cite{GHN}.
Therefore Higgs exchange effect can be larger than $W$ exchange 
by two to three orders of magnitude.

Like the $W$ contribution, charged Higgs contribution to 
$P^{\bot}_{\mu}$ also depends on the squark mixings
$|V^{D_L}_{32}|$ and $|V^{U_R}_{31}|$ (taking $|V^{H^+}_{33}|\sim 1$).
As pointed out earlier, $V^U$'s are constrained by $D\overline{D}$ mixing
only in the product of $V^{U_L,R}_{31}$ and $V^{U_L,R}_{32}$, and 
$|V^{U_R}_{31}| \sim {\cal O}(1)$ is still allowed.
On the other hand, assuming $|V^{D}_{33}| \sim {\cal O}(1)$,
the FCNC process $b \rightarrow s \gamma$ can put a bound on
$|V^{D_L}_{32}|$ from the gluino-$\tilde{b}$ diagram.  
However, other SUSY contributions to $b \rightarrow s \gamma$, 
 including the charged Higgs and chargino contributions \cite{bsgamma}, 
can dominate  over the 
gluino effect and render the bound on $|V^{D_L}_{32}|$ meaningless. 
This  is particularly true if the chargino is relatively light.
To estimate the upper limit on $P^{\bot}_{\mu}$ from Higgs exchange,
we therefore assume maximal squark mixings with 
$|V^{D_L}_{32}|=|V^{U_R}_{31}|=1/\sqrt{2}$ and 
take $m_{H^+}=100\; \mbox{GeV}$ and $\tan \beta=50$.
Setting $|\mu|=A_t=m_{\tilde{g}}$ and $I_{H^+}=1$, 
we get for the magnitude of $P^{\bot}_{\mu}$ in $K^+_{\mu3}$ decay 
\begin{eqnarray}
P^{\bot}_{\mu}|_{\bf{n}_{\mu}\cdot \bf{n}_{\nu}=0}^{K^+ \; 
\mbox{\footnotesize{decay}}} & < &
7 \times 10^{-3}, \label{eq:KHlimit}
\end{eqnarray}
where the upper bound in eq.~(\ref{eq:KHlimit}) is for
the kinematically allowed maximal value of 
$\frac{|\bf{p}_{\mu}|}{E_{\mu}} \simeq 0.90$.  
In the absence of squark family mixing, $P^{\bot}_{\mu}$
will be suppressed relative to the estimate given above 
by $\frac{m_sV^{KM}_{us}}{m_t}\sim 2 \times 10^{-4}$.

We note in passing that in the large $\tan \beta$ limit,
$P^{\bot}_l$ due to
Higgs exchange involving the $H^- \tilde{t}_L \tilde{b}^*_R$ coupling
may not be negligible in comparison to that involving the 
$H^- \tilde{t}_R \tilde{b}^*_L$ coupling considered above.
However, the upper limit given by eq.~(\ref{eq:KHlimit}) is not
 expected to be significantly modified. The same can be said for
 $D$ and $B$ semileptonic decays to be discussed below. 

  The present experimental limit on the transverse muon polarization in
$K^+_{\mu3}$ decay was obtained fifteen years ago at the BNL-AGS.
The combined value is
$P^{\bot}_{\mu}= (-1.85 \pm 3.60) \times 10^{-3}$
\cite{expmu}, and this implies 
$|P^{\bot}_{\mu}| < 0.9 \%$ at the $95\%$ confidence level.
The on-going KEK E246 experiment \cite{kuno} is aimed to  reach a sensitivity
of $9 \times 10^{-4}$. 
More recently,  studies on the BNL-AGS experiments show that 
to measure $P^{\bot}_{\mu}$ in $K^+_{\mu3}$ to an accuracy
of $\sim 10^{-4}$ can be done \cite{BNL}, and to achieve a higher precision
of $10^{-5}$ is not impossible \cite{marciano}. 
The above analysis suggests that whereas the KEK experiment is important
in testing the SUSY $T$ violation, higher precision measurements
at the AGS are required to pin down the SUSY parameter space, including
moderate squark inter-family mixings and the low $\tan \beta$ region.

  A similar analysis can be done for the semileptonic
decays of the charged  $D$ and $B$ mesons.
 However, we will concentrate on the latter in light that
$B$ factories will soon be operating.
We consider $B^+ \rightarrow \overline{D}^0 \tau^+ \nu_{\tau}$
to take advantage of the larger $\tau$ lepton mass.
The $W$ exchange (see fig.~1 with the external fermions
 properly replaced) gives rise to an effective interaction similar to
eq.~(\ref{eq:W}). 
 In the $B$ rest frame, we get 
\begin{eqnarray}
P^{\bot}_{\tau}|_{\bf{n}_{\tau}\cdot \bf{n}_{\nu}=0}^W
&  \simeq &  \frac{\alpha_s}{36 \pi} I_W^- \frac{|\bf{p_{\tau}}|} 
{E_{\tau}} \frac{m_t(A_t-\mu\cot \beta)}{m_{\tilde{g}}^3}
\frac{m_{\tau}m_B}{m_b-m_c} (1+ \frac{f^B_-(p_B-p_D)^2}{f^B_+m_B^2})
 \times \nonumber \\
 & & \;\;\;\;\;\;\;\;\;\;\;\;\;\;\;\;\;\;\;\;\; \times
\frac{Im[{V^{SKM}_{33}}^* {V^{D_L}_{33}}^* V^{U_R}_{32}
V^{KM}_{cb}]}{|V^{KM}_{cb}|^2}
\nonumber \\
& \simeq &   3 \times 10^{-5} I^-_W 
\frac{|\bf{p_{\tau}}|}{E_{\tau}}
\frac{m_t(A_t-\mu\cot \beta)} 
{m^2_{\tilde{g}}}
\frac{100 \; \mbox{GeV}}{m_{\tilde{g}}}
\frac{Im[{V^{SKM}_{33}}^* {V^{D_L}_{33}}^* V^{U_R}_{32}
V^{KM}_{cb}]}{|V^{KM}_{cb}|^2}, \label{eq:BW}
\end{eqnarray} 
where $m_{\tau}=1.78 \; \mbox{GeV}$, $m_B=5.28 \; \mbox{GeV}$, 
$m_b \simeq 4.5 \; \mbox{GeV}$ and $m_c \simeq 1.5 \; \mbox{GeV}$ 
are the masses of the 
$\tau^+$ lepton, $B^+$ meson, $b$ and $c$ quarks respectively,
 $p_B$ and $p_D$ are the four momenta of the $B$ and $D$ mesons,
$V^{KM}$ is the KM matrix,
and where $f^B_-$ and $f^B_+$ are the form factors
(c.f. eq.~(\ref{eq:Kff})).

In the heavy quark effective limit with $m_B,m_D \rightarrow \infty$,
$f^B_-/f^B_+= - (m_B-m_D)/(m_B+m_D) \simeq -0.48$ \cite{CL,HQET}.
Since $ m_{\tau}^2 \le (p_B-p_D)^2 \le (m_B-m_D)^2$ and
 thus $0.05 \le - \frac{f^B_-(p_B-p_D)^2}{f^B_+m_B^2} \le  0.20$,
we have neglected this correction in our estimate  
in eq.~(\ref{eq:BW}). 
As there exists no constraint on $V^{U_R}_{32}$ before  
we have enough $t \rightarrow c Z/\gamma$ events at the Fermilab Tevatron or
future high energy colliders,
$|V^{U_R}_{32}| \sim  {\cal O}(1)$ is currently allowed.
To estimate the largest possible effect from $W$ exchange,
 we assume the SUSY mass parameters in 
eq.~(\ref{eq:BW}) to be 100 GeV and maximal squark mixing 
with $|V^{U_R}_{32}| = 1/\sqrt{2}$.
Recall from eq.~(\ref{eq:IW-}) that $I^-_W$ can be of order 10
for $m_{\tilde{t}}$ and $m_{\tilde{b}}$ lighter than $m_{\tilde{g}}/2$.
It is thus seen that $|P^{\bot}_{\tau}|$ from  $W$-exchange 
is no bigger than a few $\times 10^{-3}$.
 
  The charged Higgs exchange contribution to 
$B^+ \rightarrow \overline{D}^0 \tau^+ \nu_{\tau}$
is dominated by the diagram involving the $H^- \tilde{t}_R
\tilde{b}^*_L$ coupling (see fig.~2 with the external lines properly
relabeled).
In the $B$ rest frame, for the outgoing $\tau^+$ and $\nu_{\tau}$ 
coming at right angle,
the $\tau$ transverse polarization is estimated as 
\begin{eqnarray}
P^{\bot}_{\tau}|_{\bf{n}_{\tau}\cdot \bf{n}_{\nu}=0}^{H^+} & \simeq & 
  \frac{\alpha_s}{3\pi} I_{H^+} \frac{\mu + A_t \cot \beta}{m_{\tilde{g}}}
\frac{m_t m_{\tau}}{m_{H^+}^2} \tan \beta  \frac{m_B}{m_b-m_c} 
\frac{|\bf{p_{\tau}}|}{E_{\tau}} (1+ \frac{f^B_-(p_B-p_D)^2}{f^B_+m_B^2})
 \times \nonumber \\
 & & \;\;\;\;\;\;\;\;\;\;\;\;\;\;\;\;\;\;\;\;\; \times
\frac{Im [{V^{H^+}_{33}}^* {V^{D_L}_{33}}^* V^{U_R}_{32} 
V^{KM}_{cb}]}{|V^{KM}_{cb}|^2}           \label{eq:BH}  \\
& \simeq &  6 \times 10^{-4} I_{H^+} 
\frac{|\bf{p_{\tau}}|}{E_{\tau}} \tan \beta 
\frac{\mu + A_t \cot \beta}{m_{\tilde{g}}}
\frac{(100 \;\mbox{GeV})^2}{m_{H^+}^2}
\frac{Im [{V^{H^+}_{33}}^* {V^{D_L}_{33}}^* V^{U_R}_{32} 
V^{KM}_{cb}]}{|V^{KM}_{cb}|^2}. \nonumber 
\end{eqnarray}
The size of $P^{\bot}_{\tau}$ from $H^+$ exchange  depends, 
among other things, on $V^{U_R}_{32}$, $\tan \beta$, 
 and the charged Higgs mass $m_{H^+}$.
Assuming $m_{H^+} \simeq 100 \; \mbox{GeV}$, 
the present limit on $\tan \beta/m_{H^+}$ \cite{GHN} allows  for
 $\tan \beta$ as large as  $\sim 50$. 
Models with maximal right-handed up-type squark mixing in the second
and third families can be motivated \cite{umix} with 
$|V^{U_R}_{32}| =  \sqrt{2}/2$ \cite{worah}.
Taking $|V^{KM}_{cb}| = 0.04$, $|V^{H^+}_{33}|=|V^{D_L}_{33}|\simeq 1$,
 and $m_{\tilde{t}}=m_{\tilde{b}}=m_{\tilde{g}}=|\mu|$, 
we have for the magnitude of $P^{\bot}_{\tau}$
in $B^+ \rightarrow \overline{D}^0 \tau^+ \nu_{\tau}$ decay
\begin{eqnarray}
P^{\bot}_{\tau}|_{\bf{n}_{\tau}\cdot \bf{n}_{\nu}=0}^{B^+ \; 
\mbox{\footnotesize{decay}}}
 & \le & 4 \times 10^{-1}, \label{eq:BHlimit}
\end{eqnarray}
where the upper limit in eq.~(\ref{eq:BHlimit})
 corresponds to the maximally allowed value
of $\frac{|\bf{p_{\tau}}|}{E_{\tau}} \simeq 0.73$.

This limit on $P^{\bot}_{\tau}$ is about two
 orders of magnitude bigger than that of
the $W$ exchange contribution, and is larger than the Higgs contribution
in the absence of squark mixing by a factor of 
$\frac{m_t}{m_b |V^{KM}_{cb}|} \simeq 10^3$.
At the proposed  $B$ factories, about $10^8$ $B$'s are collected per year,
and with that a precision of $10^{-2} - 10^{-1}$ can be achieved 
on $P^{\bot}_{\tau}$
in the decay $B^+ \rightarrow \overline{D}^0 \tau^+ \nu_{\tau}$. 
We therefore conclude that future $B$ factories have a promising
prospect to detect  transverse polarization of the $\tau$ 
due to large $\tan \beta$ and
large 2-3 family mixing in the right-handed up-type squark sector.

  We now come to the semileptonic $D$ decay, 
$D^+ \rightarrow \overline{K}^0 \mu^+ \nu$.
   For an estimate of $P^{\bot}_{\mu}$ in this decay,
we consider the potentially large 
contribution from charged Higgs exchange involving the
$H^- \tilde{t}_R {\tilde{b}_L}^*$ coupling.
In the rest frame of the $D$ meson, $P^{\bot}_{\mu}$ 
for $\bf{n}_{\mu}\cdot \bf{n}_{\nu}=0$ is given  by
\begin{eqnarray}
P^{\bot}_{\mu}|_{\bf{n}_{\mu}\cdot \bf{n}_{\nu}=0}^{H^+} & \simeq &
 -  \frac{\alpha_s}{3\pi} I_{H^+} \frac{\mu + A_t \cot \beta}{m_{\tilde{g}}}
\frac{m_t m_{\mu}}{m_{H^+}^2} \tan \beta  \frac{m_D}{m_c-m_s}
\frac{|\bf{p_{\mu}}|}{E_{\mu}} (1+ \frac{f^D_-(p_D-p_K)^2}{f^D_+m_D^2})
 \times \nonumber \\
 & & \;\;\;\;\;\;\;\;\;\;\;\;\;\;\;\;\;\;\;\;\; \times
\frac{Im [{V^{H^+}_{33}}^* {V^{D_L}_{32}}^* V^{U_R}_{32}
V^{KM}_{cs}]}{|V^{KM}_{cs}|^2}
\\
&  \simeq & - 3 \times 10^{-5} I_{H^+} \frac{|\bf{p_{\mu}}|}{E_{\mu}}
 \tan \beta \frac{\mu + A_t \cot \beta}{m_{\tilde{g}}}
\frac{(100 \;\mbox{GeV})^2}{m_{H^+}^2}
\frac{Im [{V^{H^+}_{33}}^* {V^{D_L}_{32}}^* V^{U_R}_{32}
V^{KM}_{cs}]}{|V^{KM}_{cs}|^2} \nonumber ,
\end{eqnarray}
where $m_D=1.87 \; \mbox{GeV}$ is the $D^+$ mass, $f^D_{\pm}$ are the 
decay form factors, and $p_D$ and $p_K$ are the four-momenta of the 
$D$ and $K$ mesons.
To estimate the upper limit on $|P^{\bot}_{\mu}|$,
we take $I_{H^+}=1$, $\frac{|\mu + A_t \cot \beta|}{m_{\tilde{g}}} \sim 1$, 
$m_{H^+}=100 \; \mbox{GeV}$  and $\tan \beta = 50$.
With maximal 2-3 family mixing in both the 
$\tilde{d_i}_L$ and $\tilde{u_j}_R$
sectors, $|V^{D_L}_{32}| = |V^{U_R}_{32}| =1/\sqrt{2}$, 
we find for the magnitude of $P^{\bot}_{\mu}$ in the decay $D^+ \rightarrow
\overline{K}^0 \mu^+ \nu$,
\begin{eqnarray}
P^{\bot}_{\mu}|_{\bf{n}_{\mu}\cdot \bf{n}_{\nu}=0}^{D^+ 
\; \mbox{\footnotesize{decay}}}
  & < & 7 \times 10^{-4}, \label{eq:DHlimit}
\end{eqnarray} 
where the bound corresponds to the maximal value of
$\frac{|\bf{p_{\mu}}|}{E_{\mu}} \simeq 0.99$.
This effect is too small to be seen at the proposed $\tau$-charm factory,
even though the upper limit is already larger by a factor of 
$\sim \frac{m_t}{m_cV^{KM}_{cs}} \sim 10^2$ than the 
corresponding contribution in the absence of squark family mixing. 

 It is important to notice that the separate upper limits on
$P^{\bot}_l$ in $K$, $B$ and $D$ decays given by 
eqs.~(\ref{eq:KHlimit}), (\ref{eq:BHlimit}) and (\ref{eq:DHlimit}),
may not be satisfied simultaneously as a consequence of 
the unitarity constraints on the squark mixing matrices.
These decays have different dependence on the squark family mixings,
$|P^{\bot}_{\mu}| \propto |V^{U_R}_{31} V^{D_L}_{32}|$ for $K^+$ decay,
$|P^{\bot}_{\mu}| \propto |V^{U_R}_{32} V^{D_L}_{32}|$ for $D^+$ decay,
and
$|P^{\bot}_{\tau}| \propto |V^{U_R}_{32} V^{D_L}_{33}|$ for $B^+$ decay.
Therefore, to have large mixing in the 2-3 sector of $V^{U_R}$ would preclude 
large mixing in the 1-3 sector of $V^{U_R}$, and vice versa.
However, this does not exclude the possibility of other mixing patterns
that may allow both $|V^{U_R}_{31}|$ and $|V^{U_R}_{32}|$ to be large.
It is thus fair to say that
there is no strong correlation among the different $P^{\bot}_l$'s,
and that separate experiments involving both kaon and $B$ factories
are required to probe the squark mixing matrices.

    The analysis given above has demonstrated several new features of
$T$ violation in the presence of squark family mixings. 
Generally speaking, SUSY contributions to $P^{\bot}_l$ 
are loop suppressed and
 mechanisms have to be found to overcome this suppression in order that
$P^{\bot}_l$ can be observed in the near future.
Misalignment between quark and squark mass eigenstates can lead to 
a sensitivity to the top mass for kaon, $D$ and $B$ semileptonic decays. 
If the relevant squark family mixing is not small,
this could lead to several orders of magnitude enhancement  
in $P^{\bot}_l$ relative to the case where the squark family mixing 
is absent.
We find that in SUSY models with large $\tan \beta$ and large 
squark family  mixings, charged Higgs exchange
effects can compensate the loop suppression and give values of
$P^{\bot}_{\mu}$ on the order of $10^{-3}$ in $K^+_{\mu3}$ and 
$P^{\bot}_{\tau}$ of order $10^{-1}$ in $B^+$ decay.
Both are very exciting prospects  for experimental detection 
at the on-going  KEK experiment E246 and the B factories respectively.
However, higher precision measurements for both decays are
necessary in order to cover moderate squark family mixings and
the low $\tan \beta$ region of SUSY parameter space. 
On the other hand, the transverse muon polarization in $D^+$ decay 
is found to be smaller than $10^{-3}$, which is too small 
for detection at the proposed $\tau$-charm factory.
   Although $T$ violation could also come from slepton family mixing,
these effects are expected to be less prominent than from squark
mixing because of the weak coupling suppression and 
$m_{\tau}/m_t$ suppression.

\vskip 0.2in

  We would like to thank G.~Eilam, K.~Kiers, Y.~Kriplovich and 
Y.~ Kuno for discussions, and 
 W.J.~Marciano for informative correspondence and
comments on the manuscript.
 This work is partially supported by the Natural Sciences and Engineering
Research Council of Canada.

\addtolength{\baselineskip}{-.3\baselineskip}

\setlength{\baselineskip}{\oldbaselineskip}

\end{document}